\documentclass[iop]{emulateapj}
\usepackage{graphicx}
\shorttitle{Comet Hale--Bopp at 32 AU distance}
\shortauthors{Szab\'o et al.}

\begin{document}

\title{Evidence for fresh frost layer on the bare nucleus of comet Hale--Bopp at 32 AU distance}

\author{Gyula M. Szab\'o\altaffilmark{1,2}, L\'aszl\'o L. Kiss\altaffilmark{1,2,3}, Andr\'as P\'al\altaffilmark{1,4}, Csaba Kiss\altaffilmark{1}, Kriszti\'an S\'arneczky\altaffilmark{1,2}, Attila Juh\'asz\altaffilmark{5}, Michiel R. Hogerheijde\altaffilmark{5}}
\altaffiltext{1}{MTA CSFK, Konkoly Observatory, Konkoly Thege Mikl\'os \'ut 15-17, H-1121 Budapest, Hungary} 
\altaffiltext{2}{ELTE Gothard-Lend\"ulet Research Group, H-9700 Szombathely, Hungary}
\altaffiltext{3}{Sydney Institute for Astronomy, School of Physics A28, University of Sydney, NSW 2006, Australia}
\altaffiltext{4}{Department of Astronomy, Lor\'and E\"otv\"os University, P\'azm\'any P\'eter s\'et\'any 1/A, Budapest H-1117, Hungary}
\altaffiltext{5}{Leiden Observatory, Leiden University, P.O. Box 9513, 2300 RA Leiden, The Netherlands}

\begin{abstract}
Here we report that the activity of comet Hale--Bopp ceased between late 2007 and March, 2009, at about 28~AU distance from the Sun. At that time the comet resided at a distance from the Sun that exceeded the freeze-out distance of regular comets by an order of magnitude. A Herschel Space Observatory  PACS
scan was taken in mid-2010, in the already inactive state of the nucleus.
The albedo has been found to be surprisingly large (8.1$\pm$0.9\%{}), which exceeds the value
known for any other comets. With re-reduction of archive HST images from 1995 and 1996, we confirm 
that the pre-perihelion albedo resembled that of an ordinary comet, and was smaller by a factor of two than the post-activity albedo. Our further observations with the Very Large Telescope (VLT) also confirmed that the albedo
increased significantly by the end of the activity. We explain these observations by proposing
gravitational redeposition of icy grains towards the end of the activity.  This is plausible for such a massive body in a cold environment, where gas velocity is lowered to the range of the escape
velocity. These observations also show that giant comets are not just the upscaled versions of the comets we
know but can be affected by processes that are yet to be fully identified.

\end{abstract}

\section{Introduction}

Both star formation theories and recent observations suggest that young planetary systems harbor large populations of giant comets, possibly numbering in the millions (Roberge et al. 2000, Beichman et al. 2005, Lisse et al. 2012).
These icy bodies that are more than 50 km in diameter transport water, and probably
organic matter, around their host star.  In our Solar System, C/1995 O1 
(Hale--Bopp, see Fig. 1. for an imagery) is the only giant comet (e.g. Biver et al. 1997, Weaver et al. 1997), and represent the largest and the best-studied comet in modern times of astronomy. Around its perihelion ($T_P$=April 1.19, 1997) it was visible with the naked eye, and maintained the activity level for at least 10 more years (Szab\'o et al. 2008). It is known to have been recently captured from the  Oort-cloud, possibly a few thousand years ago (Marsden, 1997) and still preserving the ancient volatile and dust content of the Solar System. 

Even when discovered, Hale--Bopp was very active and the nucleus was embedded in a thick
dust coma. The only opportunity for a photometric coma-nucleus separation was provided by 
the HST.

After the discovery, several attempts were made to determine the size of the nucleus, based on HST measurements 
(e.g. Weaver \&{} Lamy 1997; Fern\'andez 2000, Lamy et al. 2004). However, the exact size was impossible to determine 
from unresolved optical observations due to the degeneracy with the albedo. The obscured nucleus was observed at 
practically unknown rotation phases and there was no information on its shape elongation, and these had led to a 
controversy in size estimates. The most probable values for the mean radius were reported around 30$\pm$10 km, 
assuming an albedo of 4--5\%{}. The nucleus was found to rotate with a period of 11.35$\pm 0.02$ hours, which caused 
strong light variation of the inner coma (Licandro, J. et al. 1998). 

Here we present an analysis of new VLT observations, completed with archive HST data and a Herschel observation on 
the public domain. As a result of combining all these data, we derive the size, color, albedo and shape elongation of the nucleus
with a precision that was not possible before. 
Our main conclusion is that the post-perihelion albedo is significantly larger than its pre-perihelion value, and suggest a 
scenario that can explain this observation.

\section{Observations and data analysis}

\begin{figure*}
\begin{tabular}{ccccc}
1995, HST & 2009, HST & 2010, Herschel & \multicolumn{2}{c}{2011, VLT}\\
\includegraphics[height=5cm]{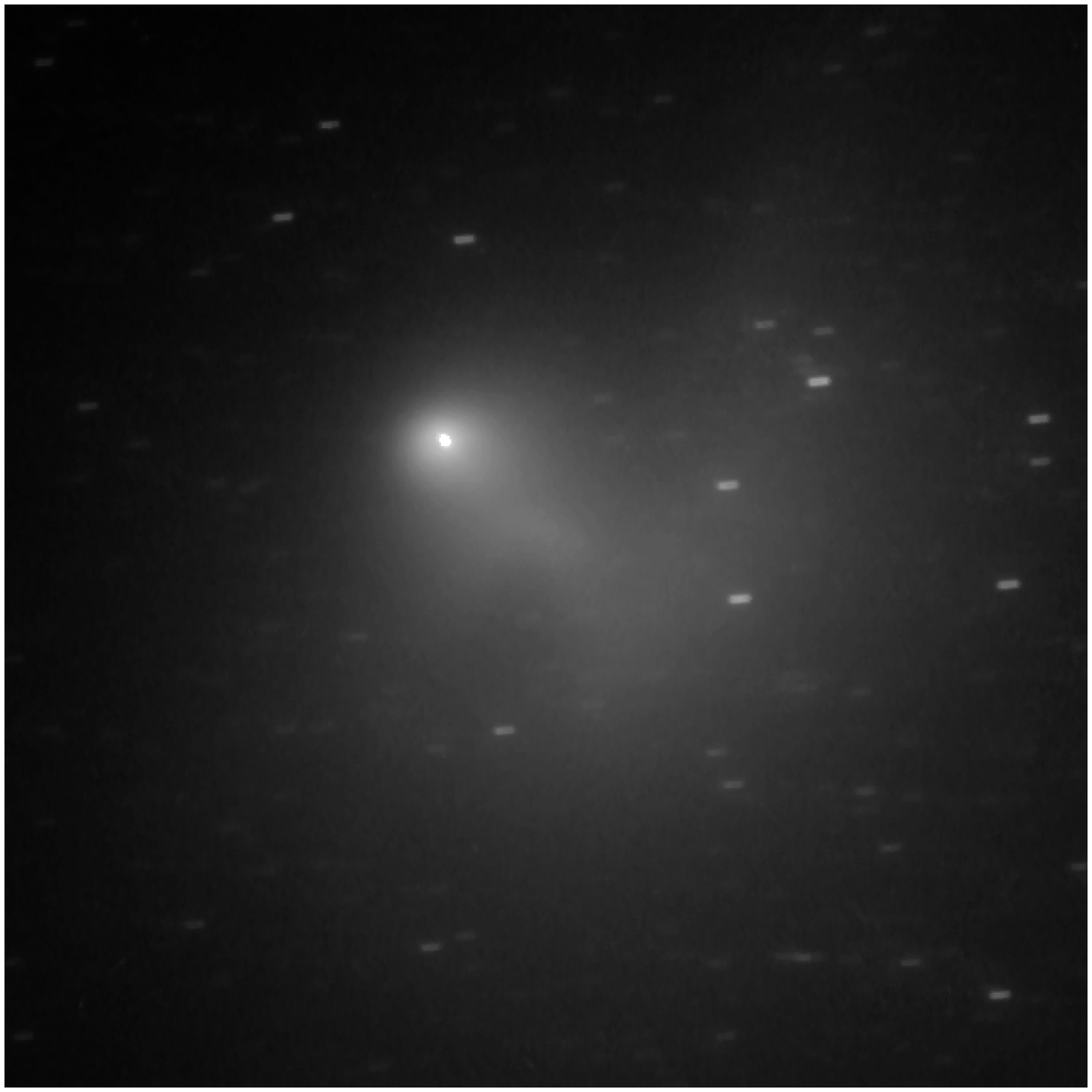}&
\includegraphics[height=5cm]{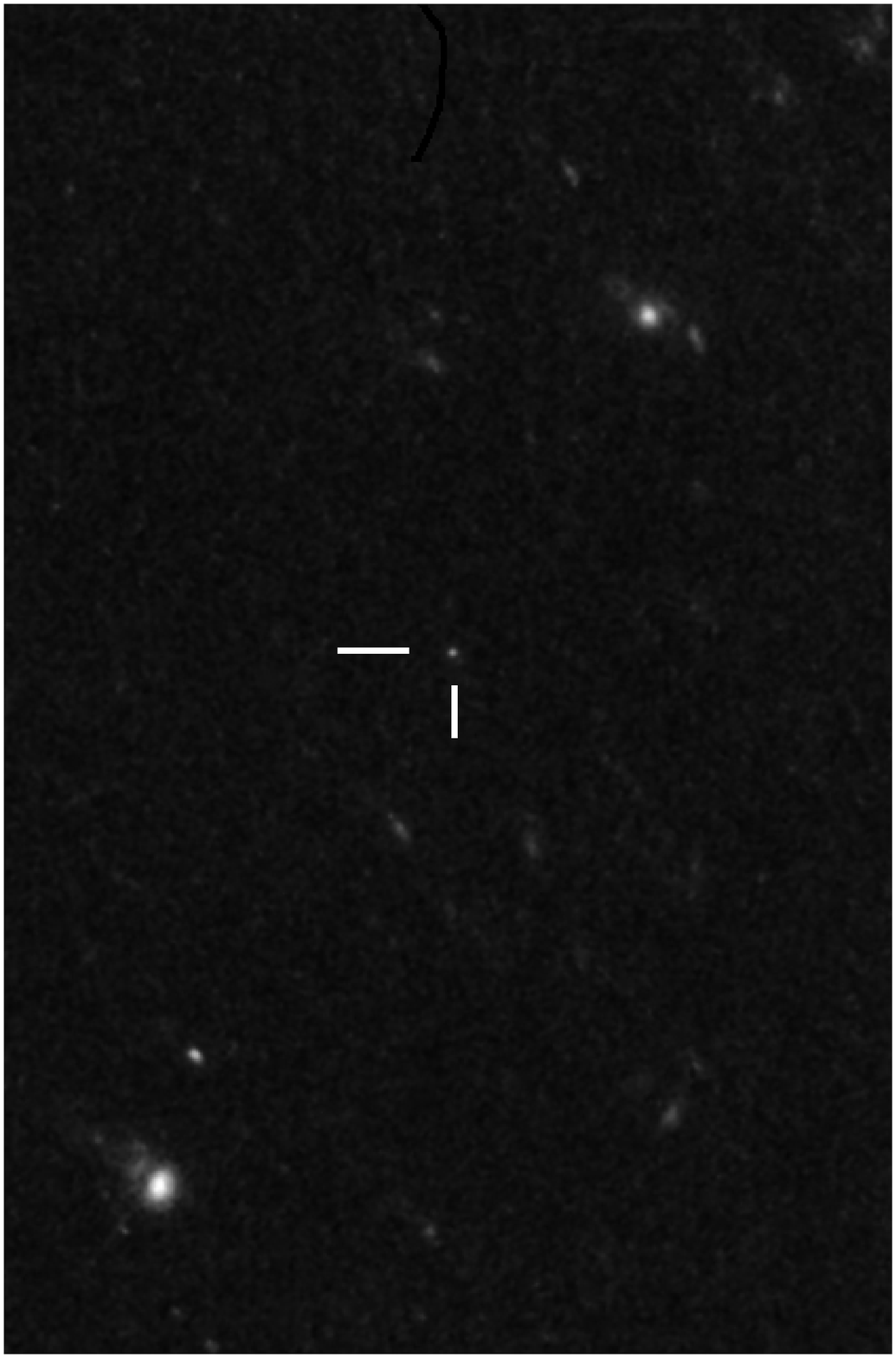} &
\includegraphics[height=5cm]{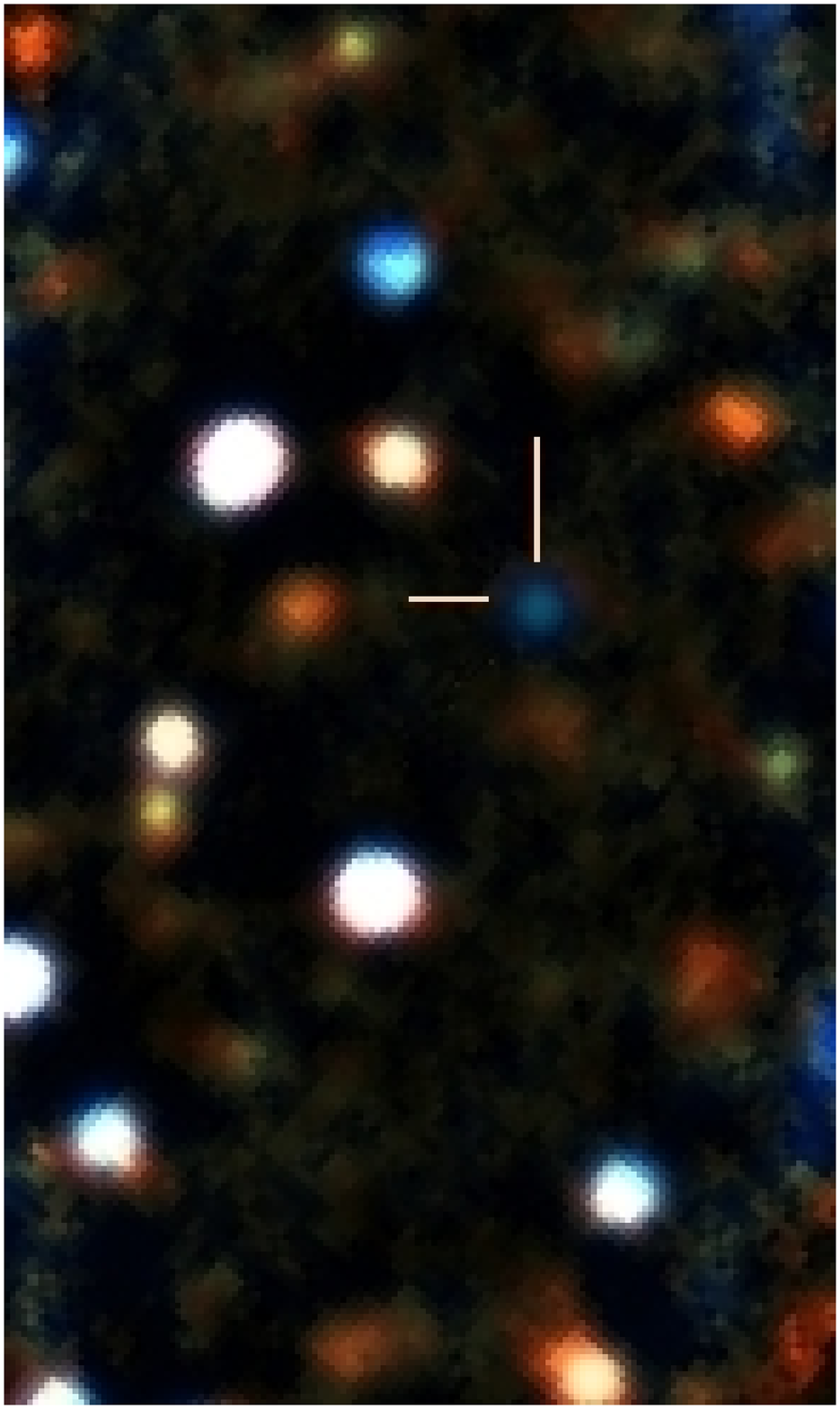} &
\includegraphics[height=5cm]{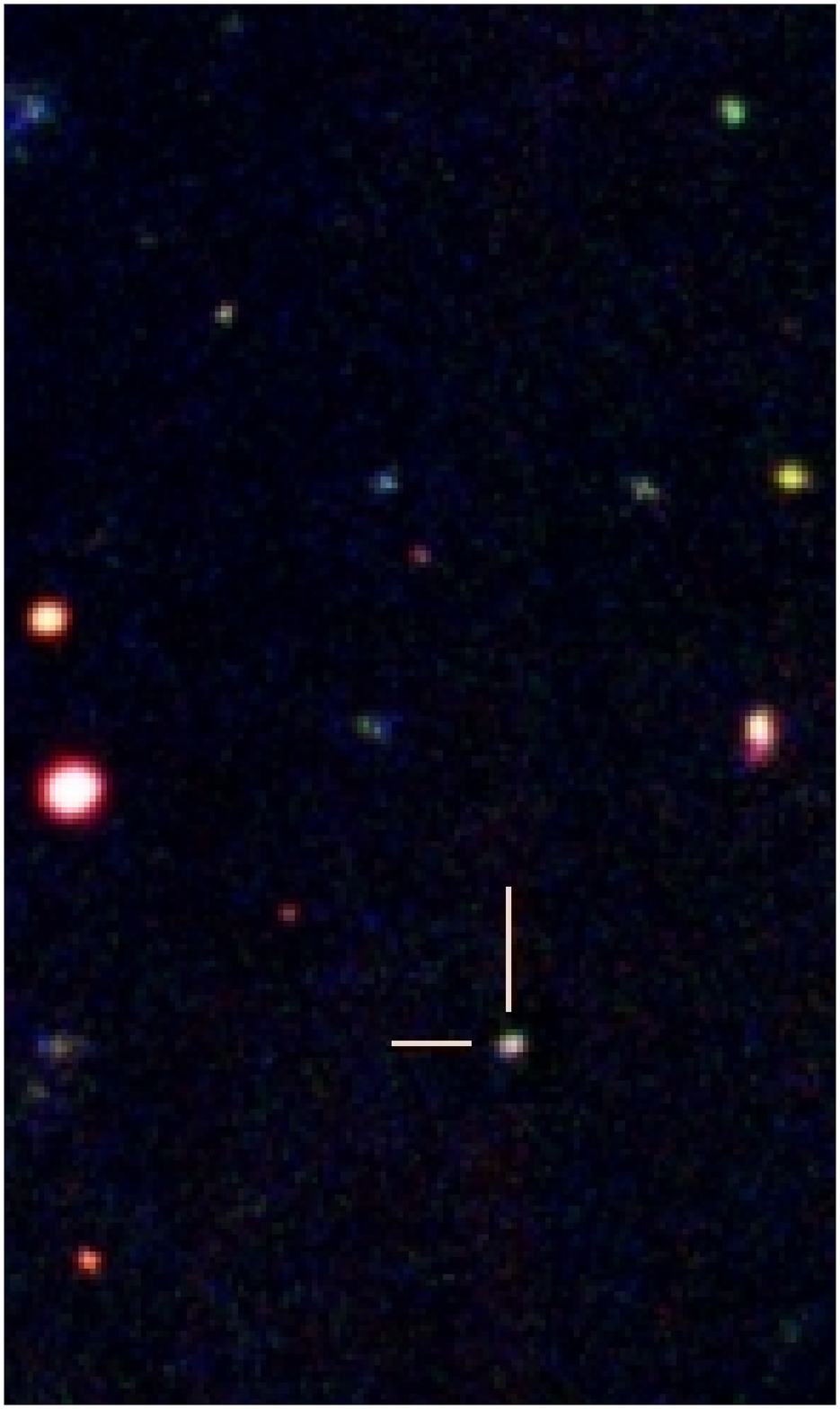} &
\includegraphics[height=5cm]{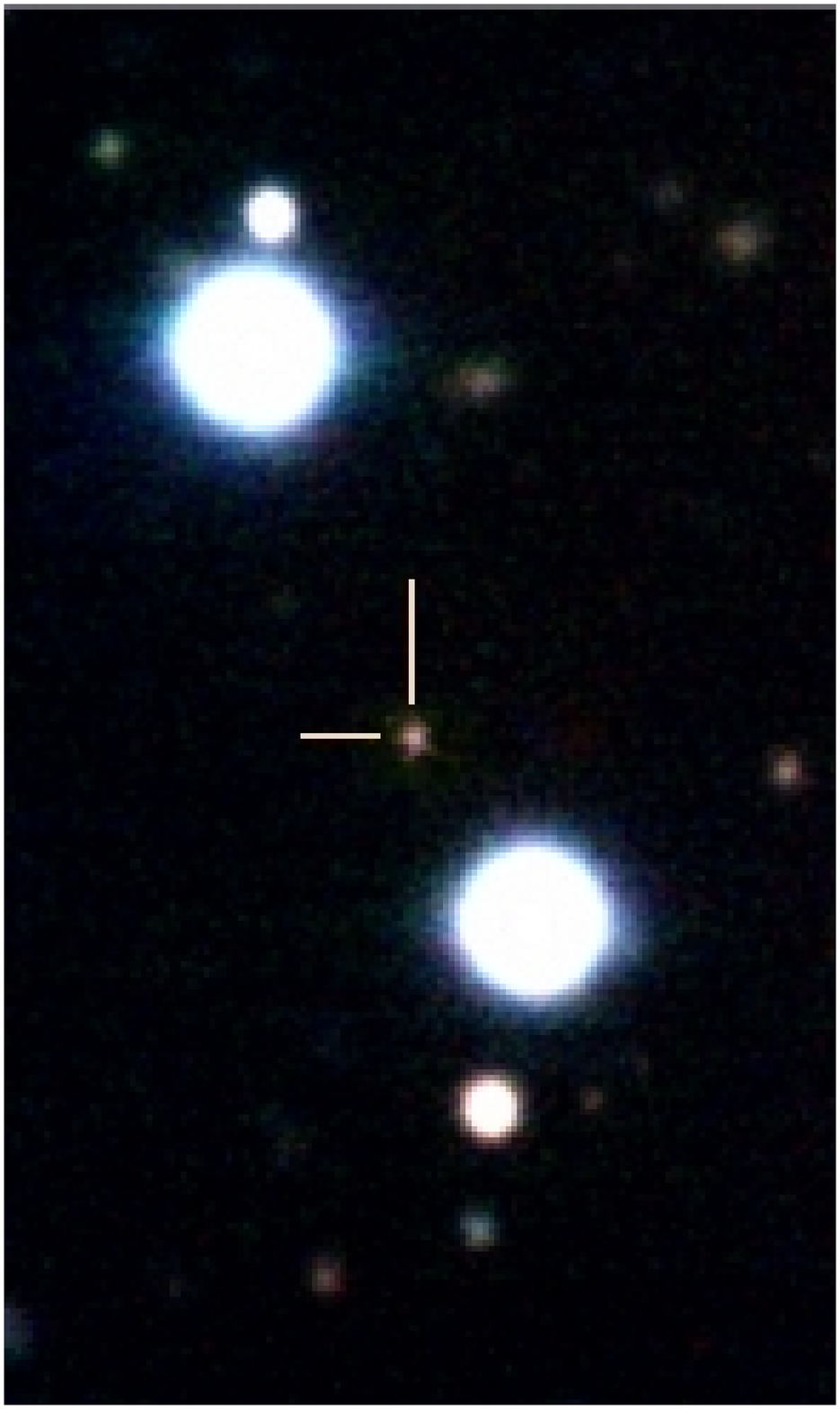}\\
\\
\end{tabular}
\caption{Imagery of Hale--Bopp. Left panel: HST ACS image in 1995 (field of view: 1.05$\times$1.05 arc minutes). Four panels to the right: post-perihelion observations with HST on 8th March 2009 (FOV: 1.00$\times$0.66 arc minutes), Herschel on 10th June 2010 (FOV: 3.00$\times$4.85 arc minutes), and VLT on 5th and 23th October, 2011 (FOV: 18$\times$31 arc seconds).}
\end{figure*}

\subsection{Observation with the Herschel space observatory}

To constrain thermal properties of the nucleus we used publicly available 
Herschel Space Observatory (Pilbratt et al., 2010) data. The observations were 
taken with the photometer of the PACS instrument (Poglitsch et al., 2010) in scan-map mode, 
on 10 June, 2010, between 00:35 and 11:10 UT in the 70\,$\mu$m and 160\,$\mu$m channels
(OBSIDs 1342198432,...1342198443). A false-color composite is depicted in Fig. 1. 
This was the longest integration made during the entire mission -- although the nucleus of Hale--Bopp is small in 
comparison to other Herschel targets, the long observation enabled us to determine the flux from the nucleus 
with a fairly good accuracy. 

Raw PACS data have been processed using a reduction script written in HIPE (Ott, 2010), 
optimized for faint solar system targets (P\'al et al., 2012a).  
For the photometry, we stacked images following the comet itself (``comet image''), and subtracted a model background 
(the combination of the observed images, without comet tracking). In order to maximize the signal-to-noise ratio, 
we selected instrumental frames using criteria based on the actual scanning speed while the frames 
themselves have been stacked after correcting the apparent proper motion of the target. 
To evaluate the averaged and background subtracted Herschel images and perform photometry, 
we used the FITSH package (P\'al, 2012b). 

PACS images yielded a flux-density of 1.08$\pm$0.16\,mJy at 70 micron while we 
obtain a 2-sigma upper limit of 0.45 mJy at 160 micron. The detected flux can be 
interpreted as the mean infrared brightness of the nucleus, because our measurements 
lasted very close to one rotation period (Licandro et al. 1998), thus averaging out 
all rotational effects. The infrared light is due to the thermal emission from the comet, 
and assuming a simple thermodynamic model of the solid body 
($\eta=1.15$ STM, Lebofsky et al. 1986), the size of the nucleus can be determined. 
Knowing the size, the optical albedo can also be calculated. We fitted the observed 
optical and infrared fluxes simultaneously, resulting in a model body with 37 km 
radius and 8.1$\pm$0.9\%{} albedo. The correlations between albedo and size are plotted in 
Fig 3 as they can be determined from optical and infrared measurements. The STM solution 
is centered at 37$\pm$3 km size and 8.1$\pm$0.9\%{} albedo. 

The reliability of the interpretation lies on how valid the STM assumptions are in the 
case of Hale--Bopp -- on the other hand, there is only one single infrared flux that we 
fit and only STM with two free parameters (size, albedo) can avoid of overfitting. 
Moreover, the PACS integration covered one full rotation of the nucleus, 
and in this case, the effect of internal heat distribution can the least modify the 
observed fluxes, so the derived size is still reliable if the STM assumptions are not 
completely fulfilled. The chosen $\eta=1.15$ parameter we applied here is the typical 
value for scattered and detached TNO population (Santos-Sanz P. et al. 2012) and we 
think it leads to a reliable size estimate because of the similarity between TNOs and 
the nucleus of Hale--Bopp. We note that $\eta$ has only a moderate effect on the size 
estimate: assuming the widely spread assumption of $\eta=0.756$ leads to a size of 
32$\pm$2 km and 12$\pm$1\%{} albedo, while assuming $\eta=1.4$ leads to 40$\pm$ 
3 km size and 7.8$\pm$0.9\%{} albedo.

\begin{table*}
\caption{Summary of Hale--Bopp nucleus photometry before and after the perihelion. Errors of last digits are given in parentheses. The last row lists albedo, assuming the radius of the body is 37 km (based on Herschel photometry). The albedo has evidently increased after perihelion.}
\renewcommand{\arraystretch}{1.2}
{\footnotesize\tt
\begin{tabular}{lrrrlcllll}
Date & R  & $\Delta$  & $\alpha$ &  & activity & \multicolumn{2}{c}{photometry}  & albedo \\
\hline
1995.10.23 &  6.36 & 6.71 & 8.1 & HST & active  & R=18.10(20) & R(1,1,$\alpha$=2)=9.55(30) & 3.6$\pm$1.0\\
1996.05.20 &  4.36 & 3.69 & 10.8 & HST & active & R=15.73(30) & R(1,1,$\alpha$=2)=9.70(40) & 3.1$\pm$1.3\\
\hline
\hline
2009.09.08 &  28.00 & 28.14 & 2.1 & HST& $>25.8$ mag/arcs$^2$ & V=23.82(11) & V(1,1,$\alpha$=2)=9.33(11) & 7.4$\pm$0.8\\
2010.06.10 & 31.45 & 31.12 & 1.8 & Herschel & inactive & b=1.08(16) mJy  & \\
 &  & &  &  &  & r<0.45 mJy  & \\
2010.12.04$^*$ &  30.68 & 31.00 & 1.7 & MPG&$>26.5$ mag/arcs$^2$  & R=23.30(20) & R(1,1,$\alpha$=2)=8.50(20) & 9.4$\pm$1.9\\
2011.10.23 &  32.00 & 32.13 & 1.8 & VLT &$>28.0$ mag/arcs$^2$ & V=24.20(10) & V(1,1,$\alpha$=2)=9.14(10) & 8.0$\pm$0.9\\ 
 &  &&                    &  & & R=23.72(10) & R(1,1,$\alpha$=2)=8.66(10) & 8.1$\pm$0.9\\ 
 &  &&                     &  & & I=23.39(10) & I(1,1,$\alpha$=2)=8.33(10) & 8.2$\pm$0.9\\ 
\hline
\end{tabular} }
\\
$^*$Data taken from: Szab\'o et al. 2010.\\
\end{table*}

\begin{figure*}
\begin{tabular}{cc}
23 Oct 1995 & 20 May 1996\\
\includegraphics[width=7cm]{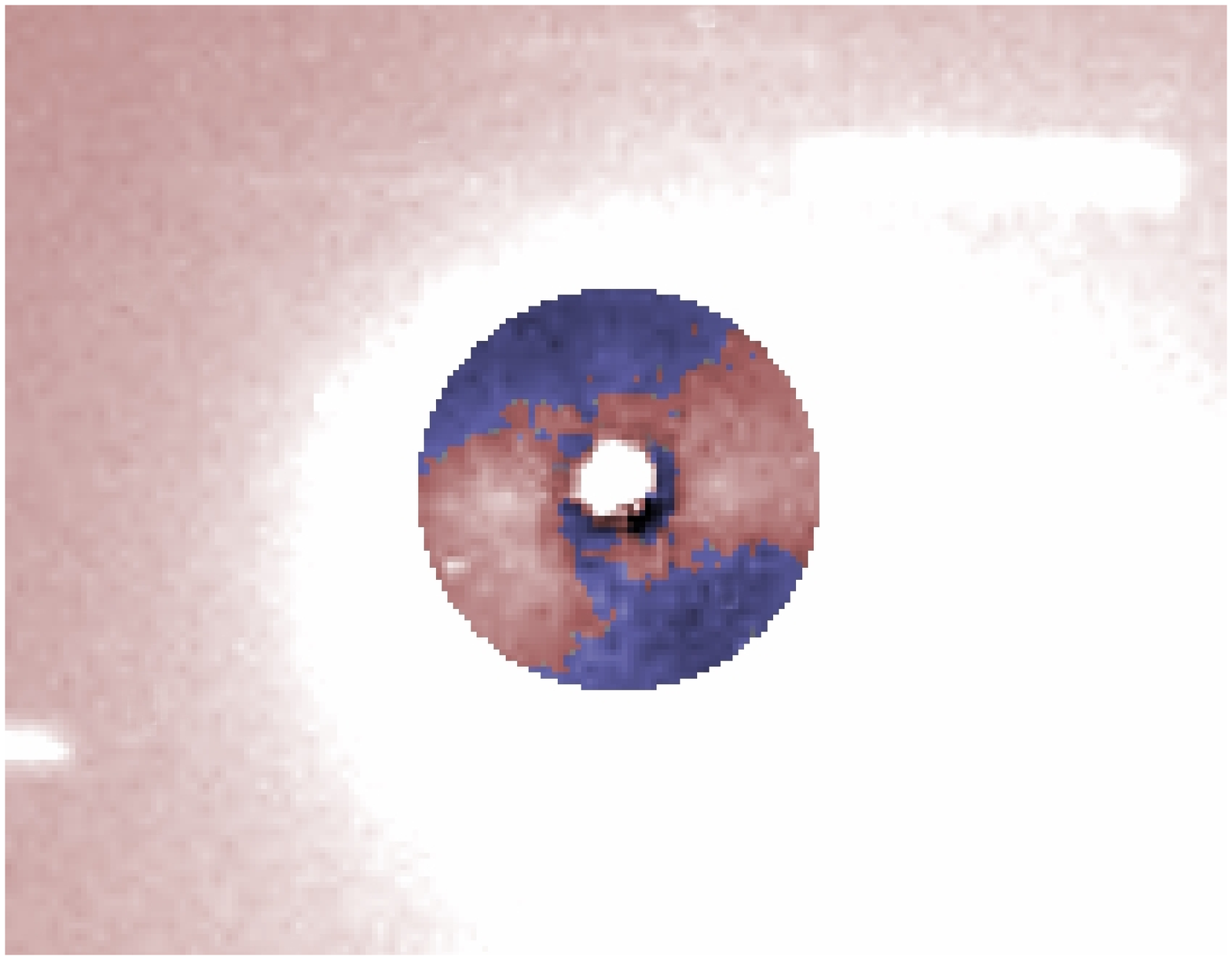}&
\includegraphics[width=7cm]{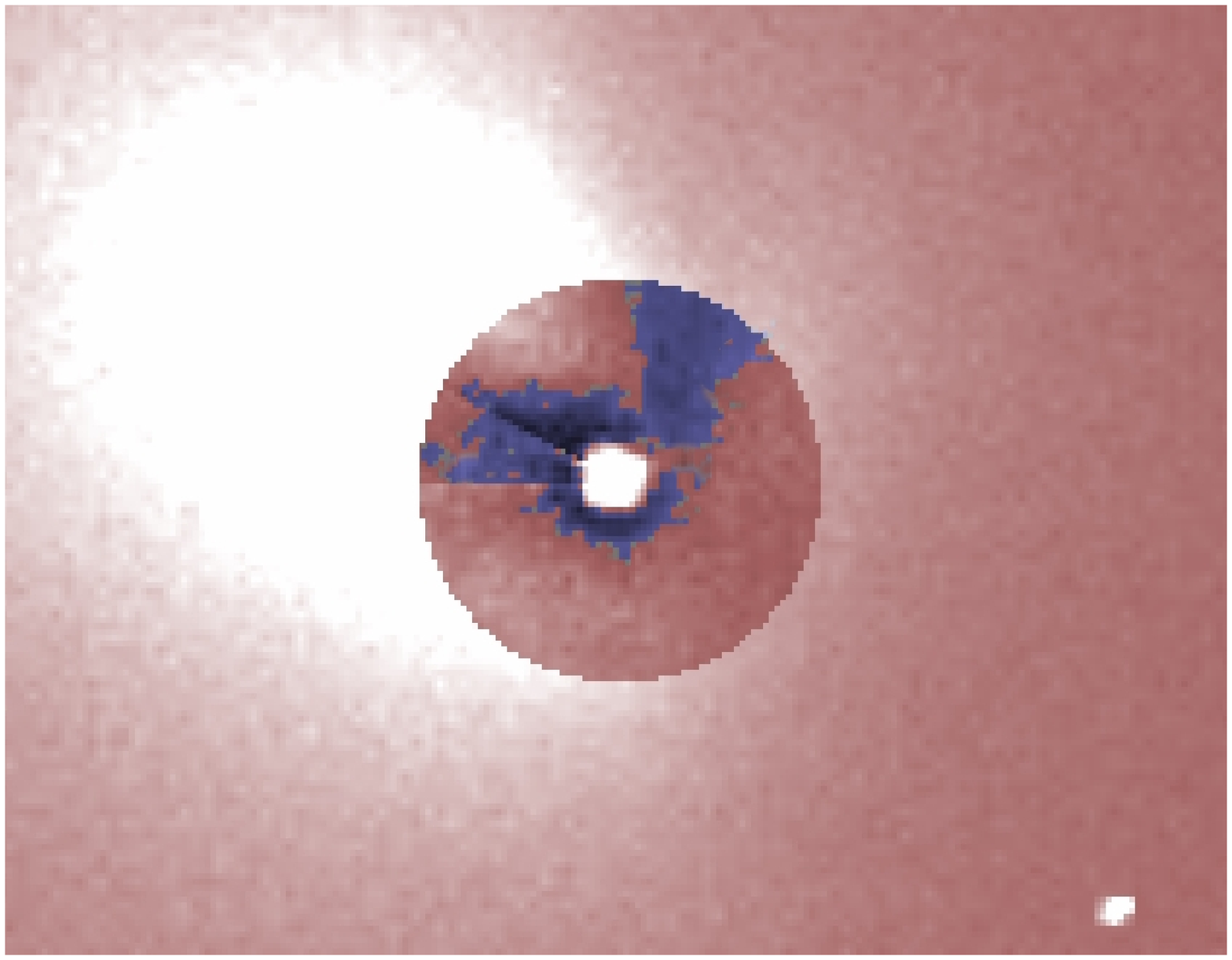}\\
\end{tabular}
\caption{Coma subtraction of pre-perihelion HST images. Positive and negative deviations are denoted by pink and blue colors, respectively. The most deviating residuals are $3$\%{} of the peak intensity in the 1995 images and $3.3$\%{} in 1996.}
\end{figure*}

\subsection{Pre-perihelion optical observations}

HST data from 1995 and 1996 were taken from the public archive (``HST Investigation of a Bright, New Comet'', 
PID 5844, PI: Harold Weaver). On 23 October 1995, four monochromatic F675W band images were taken with WFPC2, 
with exposure times of 60+300+600+600 seconds. In order to check the results, we processed another set of images 
from 20 May 1996 of the same PID (10+20+20 sec images taken with WFPC2). We checked all other exposures in the 
data archive but found a very dense coma in the case of all images taken at any other times, and their photometry 
did not lead to consistent result because of the coma contamination. So finally we accepted the photometry of images 
from 23 October 1995 and 20 May 1996.

The coma was separated from the nucleus following the recipe of image reconstruction with a power-law profile model 
coma (Lamy et al. 1998). To fit the complex internal structures of the intensity distribution of the coma component, $I(r,\alpha)$, smooth azimuthal variations were enabled in the general form of
$$
I(r,\alpha) =  {C (\alpha) \over \gamma+r^{\beta(r,\alpha)} }.
$$ 
Here $\gamma$ was a smoothing factor in the scale of the visible angular size of the nucleus (set to 0.1 pixel), 
while the log-slope of the coma, $\beta(r,\alpha)$ and the weights $C(\alpha)$ were fitted. 

Because the coma was quite complex, there might be systematics in the photometry of the nucleus introduced by 
the simplifying assumptions on the model coma. We checked the photometric stability with three different kind of coma models with various complexity. In the simplest case, both $C(\alpha)\equiv C_0$ and $\beta(r,\alpha)\equiv\beta_0$ were a constants. In the second case, azimuthal variations of $C$ were enabled with constant $\beta$, and $C(\alpha)$ was a fourth-order Fourier-polynomial of the $\alpha$ position angle. In the third case, first-order variations of $\beta$ were also allowed in the form of
$$
\beta(r,\alpha)=\beta_0 + r\beta_1 \sin (\alpha - \beta_2) .
$$
The coma was fitted out to 2$^{\prime\prime}$ distance from the center. During the fit, we applied two different optimizations as (i) the mean square of deviations and (ii) the weighted mean of square of deviations with the detected flux as weights (this second method was more sensitive in the center). Taking 3 models and 2 optimization strategies, in total 6 fits were performed for each HST observations. The photometry in WFPC2 F675W and F555W bands was corrected for aperture size and converted to $V$ and $R$ magnitudes following the calibrations of WFPC2 photometry (Holtzman et al. 1995). The parameters were fitted with an MCMC algorithm to map the confidence interval of the residual, i.e. the brightness of the nucleus.

All fitted models gave consistent results, leading to 65\%{} (23 October 1995) and 85\%{} (20 May 1996) coma contaminations. The brightness of the nucleus returned with 10\%{} standard deviation in both cases. The peak residuals of the coma were 3--4\% of the peak brightness in the original image, while the standard deviation of the residuals was 1--1.2\% around the zero mean. The favorable consistency may be due to the spatial resolution of HST: the brightness of the nucleus is coded by very few pixels, and this information can be reconstructed almost regardless how well more distant coma structures are fitted. However, we conservatively estimated 30\%{} (1995) and 40\%{} (1996) errors in the photometric results, assuming possible systematics in coma contamination that equivalently propagated to all solutions. The albedo of the nucleus was calculated assuming 37 km radius (the value determined from Herschel photometry). The quantitative results are summarized in Table 1.

From pre-perihelion HST observations we derived a pre-perihelion albedo of 2.5--4\%{}. This is consistent with several authors who got a nucleus radius around 35 km {\it if} they assumed 2.5--5\%{} albedo. Since we know {\it a priori} a reliable measurement of the size of the nucleus, we conclude that the derived albedo is consistent with the previous results.

One possible doubt about the low albedo could be raised based on the known elongation of the nucleus.
If both the 1995 and 1996 images showed the nucleus in a rotational phase when it was fainter than the average, then the albedo is underestimated. This scenario is unlikely, because the probability that both measurements show a sub-average brightness is only 25\%{}. However if this happened, the albedo should be corrected by the variation of the projected area of the nucleus ($\pm 30$\%{}), leading to an upper albedo estimate of $\approx$4.5$\pm$1.1\%{} -- still significantly darker than the post-perihelion estimates. 

The conclusion on the albedo variation is completely independent on how well the nucleus can be described with an STM model, and relies only on the measured optical brightnesses of the nucleus. This is because the same nucleus size was assumed for all albedo determinations, and the derived albedo values are all scaled by this parameter. Should the size of the nucleus be misestimated by a factor, all albedo values should be corrected by this factor simultaneously, and the conclusion on an albedo variation with a factor of 2--2.5 will be still valid.

\begin{figure*}
\hskip2cm\includegraphics[bb=180 307 432 487,width=7cm]{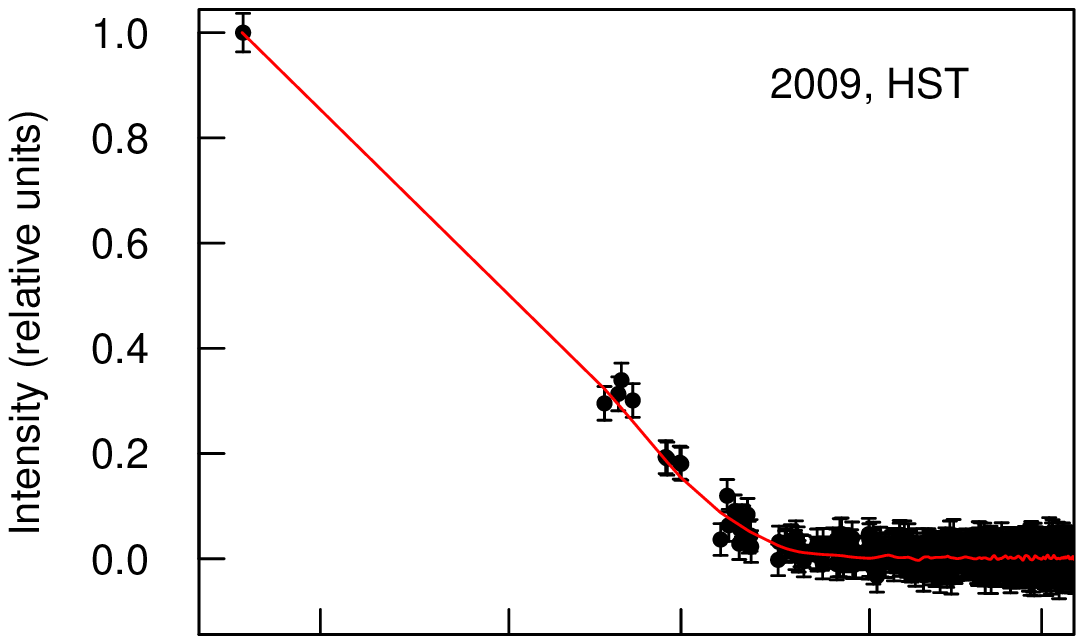}
\hskip2cm\includegraphics[bb=180 307 432 487,width=7cm]{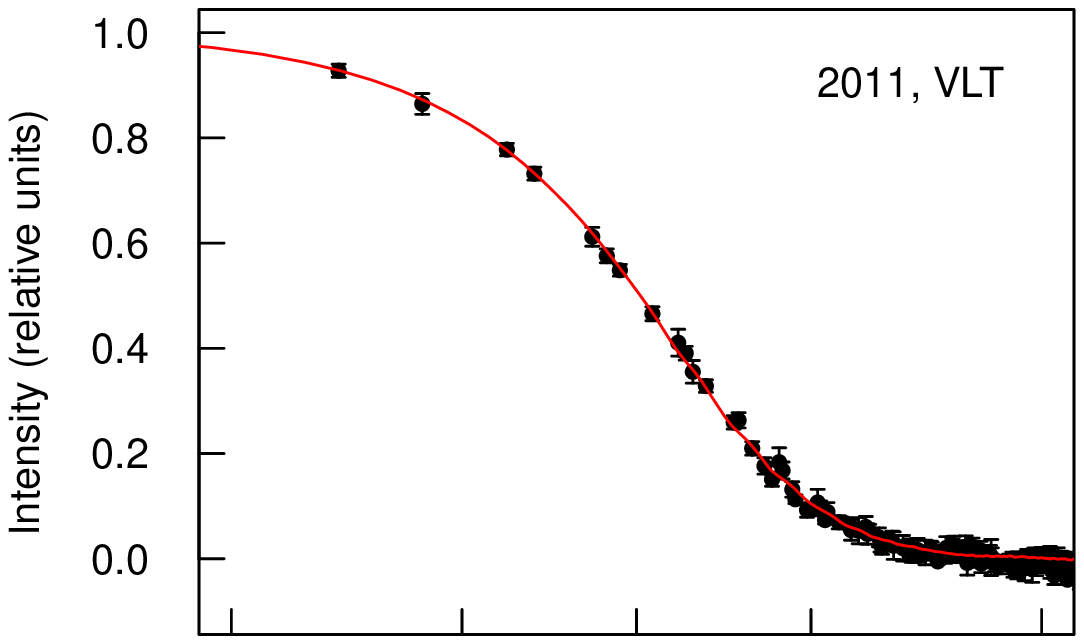}

\hskip2cm\includegraphics[bb=180 303 432 433,width=7cm]{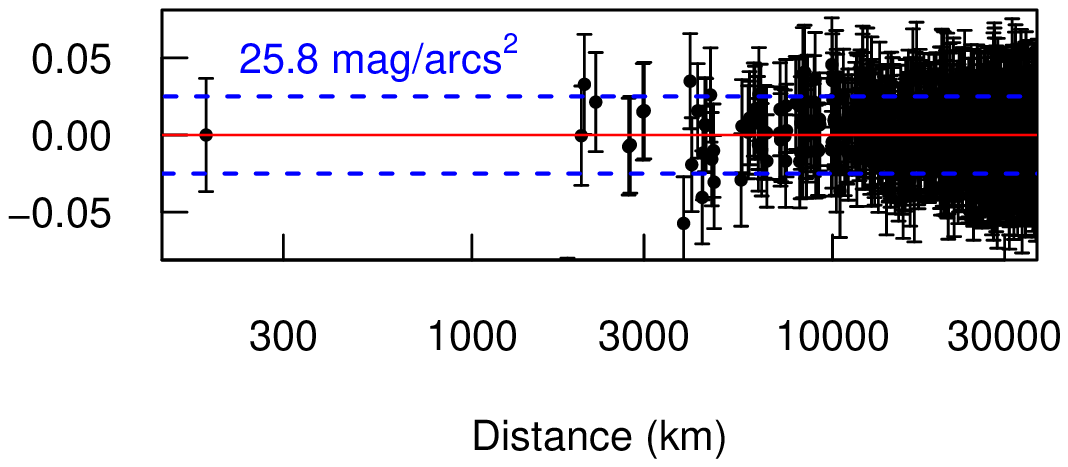}
\hskip2cm\includegraphics[bb=180 303 432 433,width=7cm]{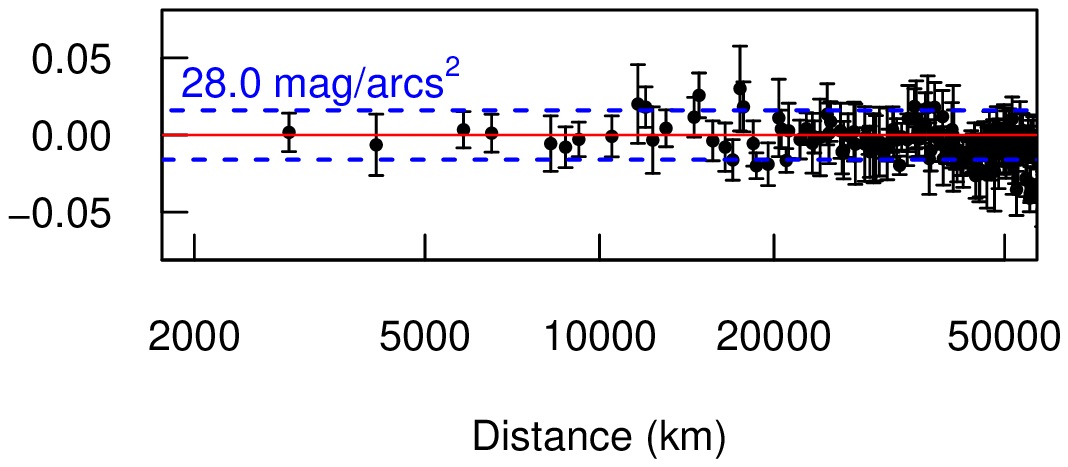}
\caption{Confirmation of the star-like appearance of Hale--Bopp after 2009. Imaged intensity profile of Hale-Bopp in 2009 (HST, F606W band, 0.5 hour integration) and in 2011 (VLT $V+R+I$ fluxes, 6 hour integration) compared to a mean stellar profile (red line). The residuals (comet$-$stellar profile) exclude a coma brighter than $\approx$25.8 and 28.0 magnitudes/arcsec$^2$ in 2009 and 2011, respectively.}
\end{figure*}

\subsection{Post-activity HST and VLT observations}

The first observation where the nucleus is bare is an archive HST image from 2009. The lack of any coma brighter than 25.8 mag/arcsec$^2$ was indicated by the observed radial intensity distribution of the comet, which was exactly compatible with a stellar image. The photometry of HST images was consistent with an increased albedo (see Table 1). We reduced the nucleus image according to the standard recipe of HST photometry, in an aperture of 0.20 arc seconds and applying the appropriate aperture correction and we converted F555W fluxes to Johnson $V$ assuming solar colors ($V-F555W=0.01$).

In October 2011, we made a deep (6 hours integration with VLT) follow-up and excluded any extended coma feature exceeding a surface brightness level of $>28$ mag/arcsec$^2$ ($<$50km$^2$ scattering cross section in a 23000x23000 km column). 
We made use of FORS2 instrument on the 8.2m Antu unit of the Very Large Telescope, located at the  Cerro Paranal site of the European Southern Observatory (Chile). Six hours of integration (1+2+3 hours centered on 2011 October 5.55, 23.63 and 25.59 UT, respectively) were obtained in V, R and I filters. The exposure times were set to 3--4 minutes to prevent trailing of the slowly moving image of the comet (sample images are shown in Fig. 1). VLT images (image scale of 0.252 arcsec/pixel, dithering applied during acquisition) were resampled to 0.126 arcsec/pixel scale and were aligned to the stellar field (geotran task in IRAF). These images were co-added to the stellar image. Then all individual images were shifted to compensate the apparent proper motion of the comet (taken from NASA/HORIZONS system), and they were co-added to the comet image.nBecause of the motion of the comet, the coma was blurred to less than the extention of a pixel (sidereal tracking was applied). Therefore a motion blur was applied to the stellar image as well, and the coma profile was compared to this trailed stellar profile. We found a perfect fit, i.e. the nucleus was star-like in this deep image.

The surface brightness of the background, 28 mag/arcsec$^2$ is compatible to the surface brightness of a homogeneous coma with less than 1 dust particle per 250 m$^3$ volume (assuming a coma diameter of 23000 km ($\equiv$ 1 arcsec in our images) with uniform dust density, 1$\mu$m dust grain size and 5\%{} albedo). Since we did not detect signs of a coma, we conclude that Hale--Bopp was surely inactive at the time of this observation. This enables us to determine the properties of the nucleus free of coma contamination for the first time. For this task, perture photometry was applied with an aperture size of 7 pixels. On all nights, two Stetson standard fields were observed at 1--1.1 and at $\approx$1.8 airmasses. Because of the slow motion of the comet, the same local comparison stars could be used for the October 23/24 and the October 25/26 observations. For a consistency check of the photometry, USNO-1B stars were tested. The averaged R2 brightness of USNO stars in the field was compatible within an error of 0.04 (Oct. 23) and 0.03 (Oct. 25) magnitudes.

From VLT and HST observations we consistently determined the absolute brightness of the nucleus as $V(1,1,\alpha=2)\approx9.2\pm0.1$ and $R(1,1,\alpha=2)\approx8.6\pm0.1$ magnitudes. This proves that the comet was already inactive in 2010, when the Herschel observations and our optical follow-up (Szab\'o et al. 2010) were performed. We measured color indices of $V-R=0.48\pm 0.10$, $V-I=0.81\pm0.10$, slightly redder than the Sun (0.36 and 0.71, respectively). Thus, the normalized reflectivity gradient is $4\pm4$\%{}, a quite low value for a comet, indicating a significantly bluer nucleus than the inner dust coma observed at 25.7 AU distance ($V-R=0.66$, Szab\'o et al. 2008). 

\begin{figure}[h]
\hskip1.2cm\includegraphics[bb=143 320 526 472,width=7cm]{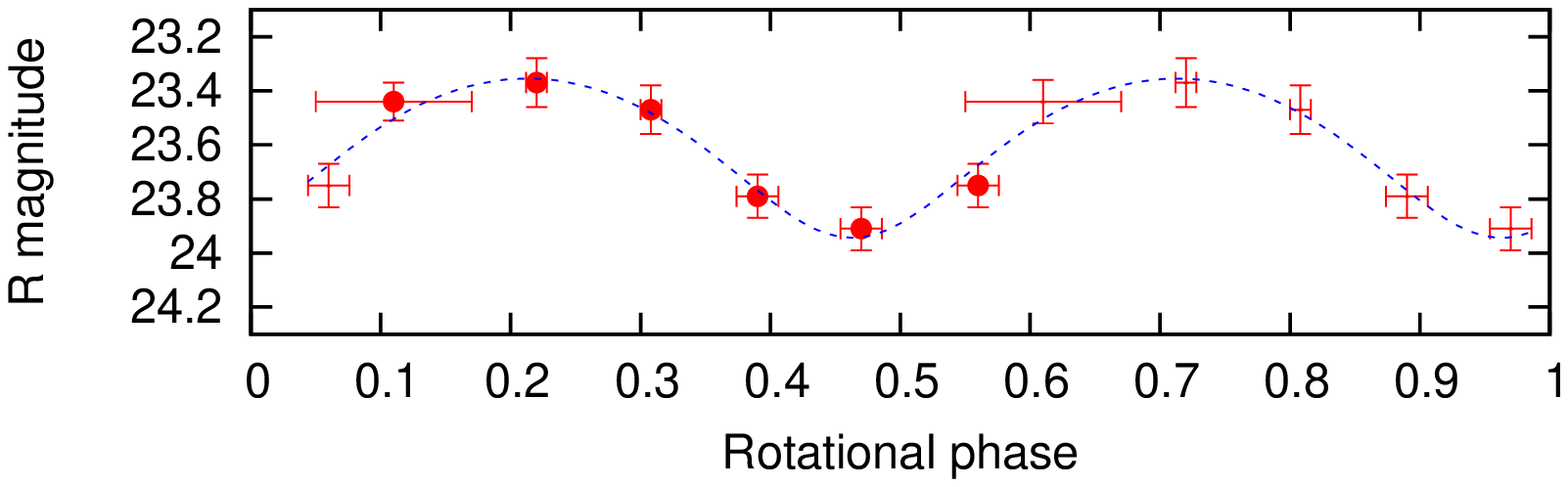}

\hskip1.2cm\includegraphics[bb=108 90 552 423, width=7cm]{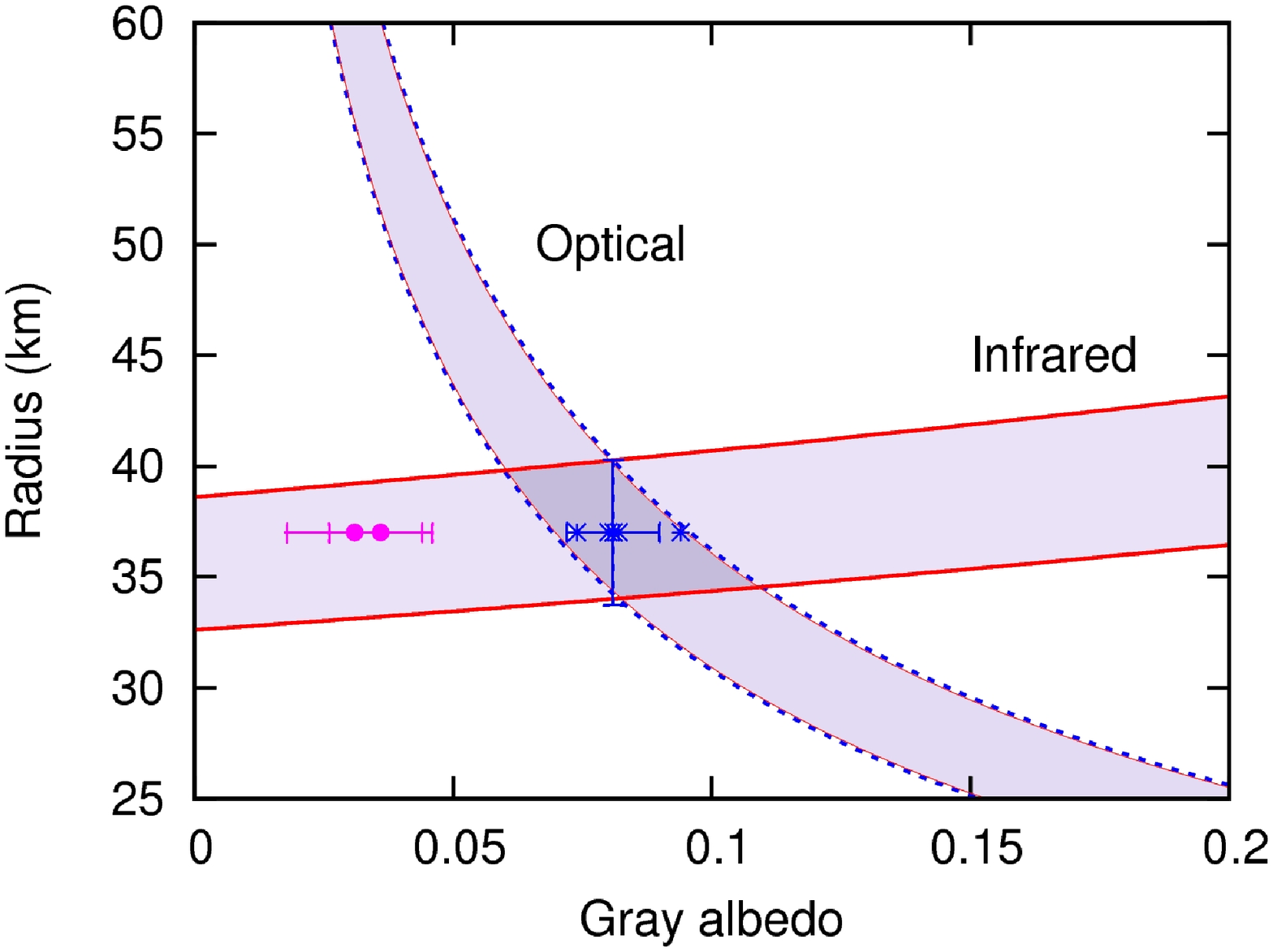}

\vskip1cm

\caption{Determination of shape elongation from time-resolved optical photometry (upper panel), and mean size  
and albedo from optical and infrared fluxes (lower panel). Data for the pre-perihelion nucleus are plotted in pink, while the frozen-out state is represented by the blue symbols.
}
\end{figure}

During the six hours of VLT photometry we also revealed variations in the apparent brightness that clearly exceed the measurement errors. Phased with the 11.35 h period and assuming identical variation for the two halves of the rotation, the curve can be fitted by that of a rotating prolate shape with $a/b\geq1.72\pm0.07$ axis ratio (equality in case of a perpendicular aspect to the rotational axis; Fig. 3). Similar shape elongations were observed by spacecrafts at 1P/Halley ($a/b=2.0$), 17P/Borrelly ($a/b=2.5$, Hsieh et al. 2009) and 103P/Hartley 2 (3.38, A'Hearn et al. 2011), all exhibiting a peanut-like or dumbell-like nucleus. The inferred shape elongation is also consistent with previously reported large-amplitude periodic brightness variations in the inner coma, which was a well-known property of Hale-Bopp in active stage. These variations were interpreted as the rotation of the nucleus, active areas and the the related coma inhomogeneities. Elongated shape amplifies near-nuclear jets and leads to severe anisotropy of the inner coma (Cifro et al. 2002), well explaining the complex coma structures that were observed in the active stage of Hale--Bopp.

\begin{figure}
\includegraphics[width=8cm,bb= 70 207 465 500]{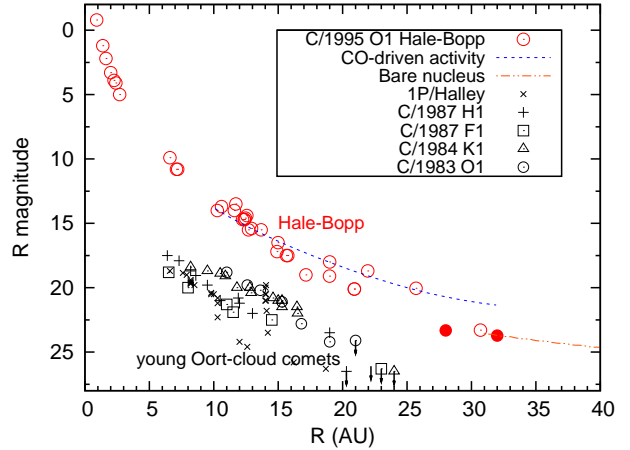}
\caption{Total brightness of comet Hale--Bopp after the perihelion passage. The blue line represents a model with
CO-driven matter production (Capria et al. 2002), the red line is the light expected from the bare nucleus. For comparison,
the typical light variation of young Oort-cloud comets is shown.}
\end{figure}

\subsection{A 25 magnitude apparent brightness variation in the history of Hale--Bopp}

In Fig. 5 we plot the total brightness estimates (both visual and CCD) of post-perihelion Hale--Bopp. The data were collected from the ICQ database and from Szab\'o et al. (2008, 2011) and references therein. Previously published brightness estimates are plotted with open circles, the new post-perihelion measurements are denoted by solid dots. Near perihelion, the comet was slightly brighter than 0 magnitude, and now it has faded to $V\approx24.2$. Thus, its observed apparent brightness variation spans 25 magnitudes, which is unprecedented for astronomical objects. 

The post-perihelion evolution of Hale--Bopp can be divided into two of three major distinct phases, as can be seen in Fig. 5. The most prominent feature is the cessation of activity between late 2007 and early 2009, somewhere between 26--28 AU solar distance. In the active phase closer than $\approx$ 15 AU to the Sun, the brightness could be well fitted with a prediction assuming CO-driven activity (Capria et al. 2002). In between, some outliers appear between 15--25 AU, which are about 1--1.5 magnitude fainter than the prediction or other measurement that fit well the prediction -- while there are no too bright outliers. This may signify that the activity was impedent near cessation, and matter production might have erratically flickering between full and exhausted levels.

\section{Discussion}

The reliability of our result on the increased albedo depends on how confidently we can exclude a significant coma contamination in the HST (2009) and VLT (2011) observations. To place an upper estimate on a possible contamination in the photometric aperture, we calculated model comet images, consisting of a nucleus and a coma component with increasing brightness in subsequent models. To sample the images properly, model comet images were projected into an image matrix with the same resolution as the actual images, and into the same subpixel position. The model pixel intensities were calculated by numerical integration in each pixels, with 10$\times$10 subpixel resolution. Brighness profiles were extracted with the same tool as for the observations, and they were similarly normalized to peak intensity. We copied the error bars from the observed image and assigned to the model pixel by pixel. This process led to identical sampling and identical error distribution as in the case of the real observations, thus the results could have been directly compared to stellar profiles (see Fig. 6 for an illustration).

\begin{figure*}
\hskip2cm\includegraphics[bb=180 247 432 487,width=7cm]{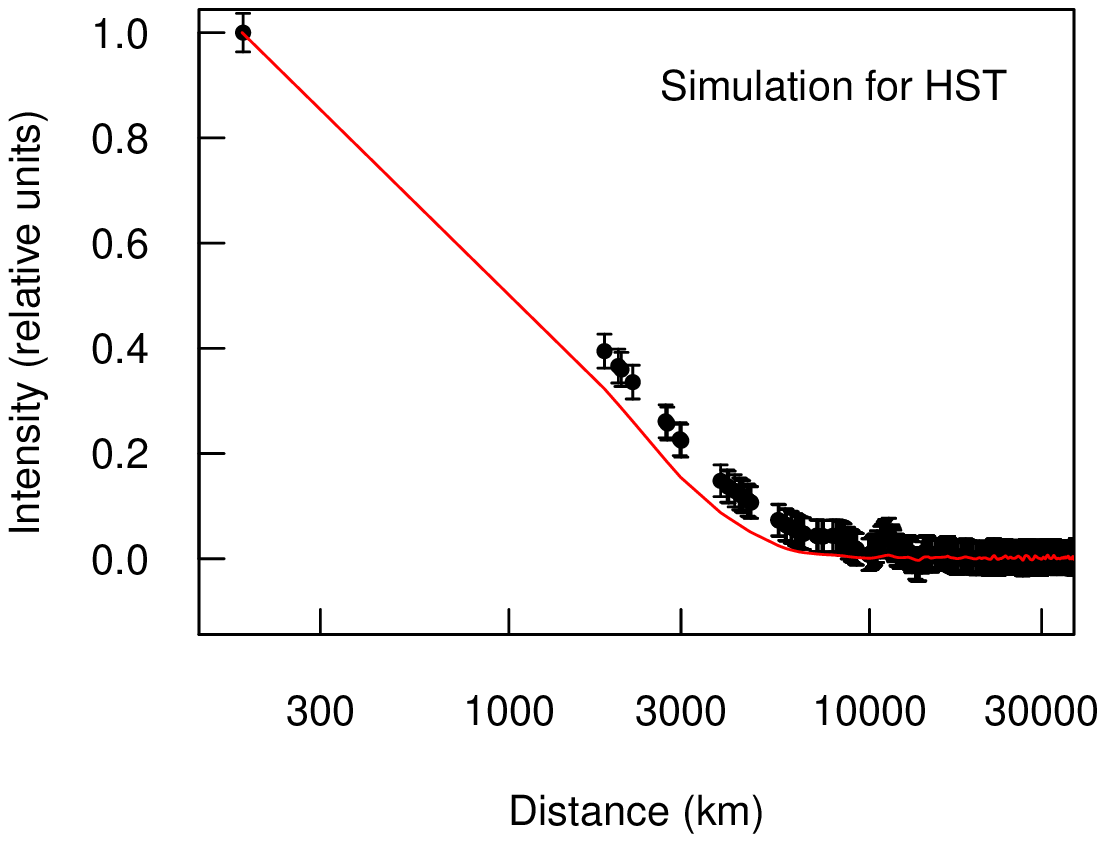}
\hskip2cm\includegraphics[bb=180 247 432 487,width=7cm]{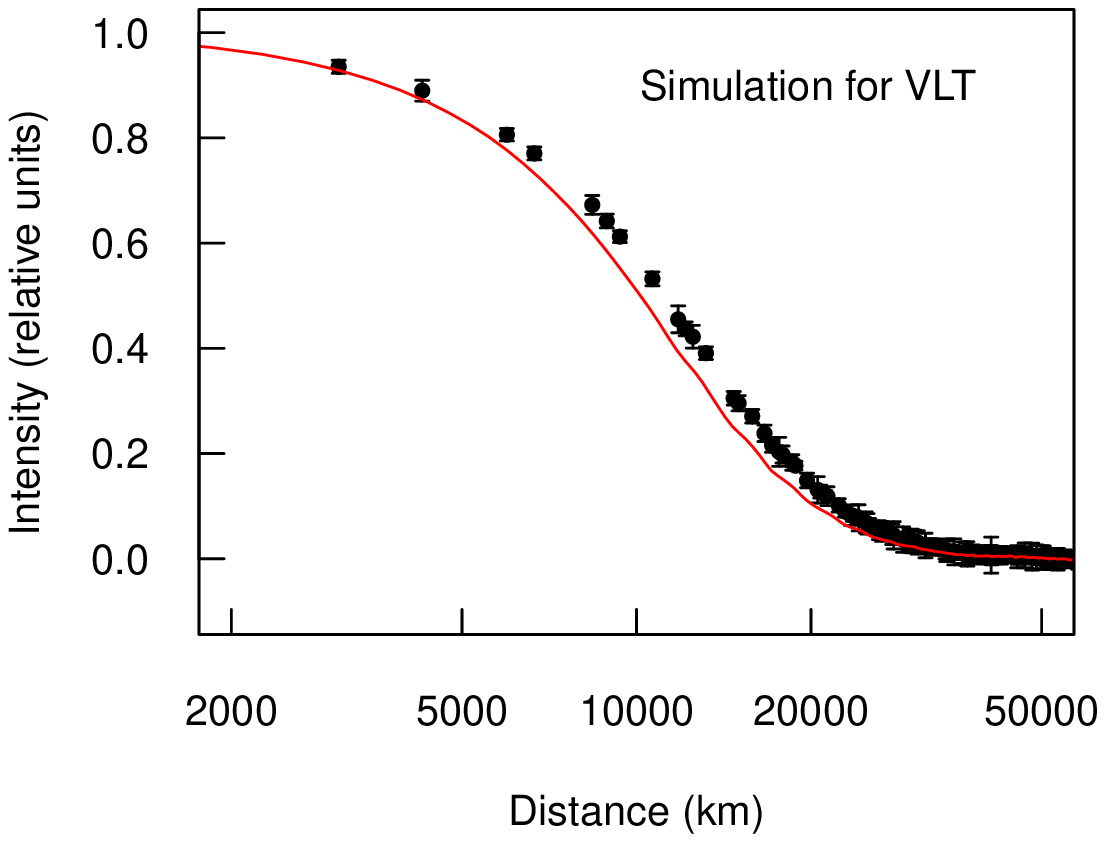}
\caption{Brightness distributions of a simulated nucleus and a faint $1/r$-law coma with 30\%{} brightness contamination in the photometric aperture, sampled accordingly to the HST (left panel) and the VLT (right) observations. Error bars are taken from the actual images. Such bright comae could have been easily detected in the observations. Undetected comae may exhibit less than 13--21\%{} contamination, depending on the assumed shape of the coma model.}
\end{figure*}

The largest possible coma contamination slightly depends on the coma structure (Jewitt and Danielson 1984): the more compact the coma the more it boosts deviations near the photocenter -- therefore significant deviations from the stellar profile may emerge at a lower contamination from a compact coma than for a loose coma with higher total contamination. Because the structure of the hypothetical coma is unknown, we tried three different model comae in our analysis. A loose coma was represented by a brightness distribution of $\rho^{-1}$ (where $\rho$ is the distance from the nucleus in celestial projection), which is typical for comets near perihelion. At large solar distances, comets have more compact comae, typically $\rho^{-2}$, which was the brightness distribution of our second test coma. In 2007, Hale-Bopp exhibited a coma with very flat brightness distribution, that was almost constant in the inner 50,000 km and then suddenly disappeared. Therefore, our third coma was a staircase function of the radius, exhibiting a constant plateau in the inner 2$^{\prime\prime}$ and being zero elsewhere. These model comae were convolved with the stellar PSF and added to the model nucleus image. The smallest significant deviation was determined with observing the $\chi^2$ between the model coma and the stellar PSF, aiming at a 4-$\sigma$ level. 

The model image with 4-$\sigma$ deviation then was re-normalized to the central intensity of the nucleus component only, to recover the brightness of the coma component properly. Photometry was performed in the same aperture as in the case of the real observations. The relative flux of the model (nucleus$+$coma), compared to the photometry of the model nucleus alone, revealed the largest contamination from an undetected coma.

Such way, the combined HST image was derived to be contaminated by less than 13\%{} (staircase coma model), 18\%{} ($\rho^{-2}$ model) or 20\%{} ($\rho^{-1}$ model.) Similarly, the combined VLT observations are contaminated by less than 15\%{}, 19\%{} and 21\%{}, respectively. (The similarities are only accidental: the HST images have much better resolution, while the VLT images are much deeper.) One may conclude that once the activity level suddenly decreased significantly and then apparently reached a constant absolute brightness level (as measured in 2009, 2010, and 2011), the comet has reached its inactive state. 
However, based on our simulations, we clearly find that even if there was a combination of a nucleus and a faint unresolved coma, the photometric contamination was certainly less than $\approx$20\%{}, both in 2009 and in 2011, and therefore, the derived increment of the absolute brightness does indeed reflect an increment of the surface albedo.

\section{Conclusions}

The derived $\sim$8\% albedo is evidently larger than those known for any other comets. To check the pre-perihelion value, we reprocessed archive HST observations from 1995 and 1996; separated the nucleus from the coma; measured its absolute brightness and derived the albedo, assuming the inferred 37 km mean radius.
The pre-perihelion albedo was calculated to be 3$\pm$1\%, where the quoted uncertainty comes from the residuals of the imperfect coma subtraction. However, it can be safely conclude that the albedo was significantly lower before perihelion than thereafter: even without subtracting a coma model from the early HST data, the brightness of the photocenter was as bright as a single body would have been with 37 km in radius and 8-9\% in albedo. Because of the evidently strong contamination by the thick inner coma, the nucleus must have been much darker. 

The sudden change of albedo reflects a significant change of surface properties post-perihelion. This could happen at the final phase of matter production by ballistic redeposition (Jewitt 2002) of the ejected matter. For an order-of-magnitude estimate of the redeposition rate, we determined the escape velocity on the nucleus which is $v_e=10$--$13$m/s (on the ends and waist, respectively; assuming a bulk density of 1000 kg/m$^3$), and the thermal velocity of sublimating CO gas at 32 AU at radial direction, which is $v_{th}=88$m/s. Because these velocities are of the same order of magnitude, a significant fraction of the dragged matter can fall back to the surface. Following the standard recipe (Jewitt 2002), the capture fraction is estimated to be 2.6\%{} on the ends and 3.5\%{} on the waists of the nucleus. Due to an activity level of 250 kg/s as observed in 2007, the rate of mantle formation is 0.16 mm/day. This redeposition rate is enough to entirely change the optical properties of the nucleus in a few days. Hale--Bopp has been the first example for such significant resurfacing by ballistic redeposition -- because the nucleus of other comets is smaller and the activity ceases in a warmer environment, the redeposition rate is insignificant in their case.

The pre-perihelion albedo of the nucleus, like in the case of all known cometary nuclei and also D-type asteroids, was caused by the presence of processed organic matter on the surface. After the activity had ceased, there remained a layer of redeposited grains. To support the increased albedo, this layer must have lacked organic matter and contain silicates and/or fresh ice. Silicate-rich surfaces are seen on S-type asteroids in the inner asteroid belts (indicatively at $<$2.2 AU solar distance), which have typical albedo in the range of 10--20\%{}. These asteroids suffer permanent irradiation, they have entirely lost their volatile content, and evidently look very different from long-period comets. Moreover, S-type asteroids have typically slightly negative $R-I$ spectral slope (Solontoi et al. 2010), in contrast to the neutral color of Hale--Bopp's nucleus. 

Both the increased albedo and a neutral $R-I$ color can be supported by significant amount of icy grains on the surface. These ice grains can not sublimate in the cold environment of $\approx$50 K, and will be preserved for a long time. The observed albedo of 8\%{} resembles the value of Centaurs, which are small bodies orbiting among the outer giant planets and have either fresh water ice on surface (e.g. Asbolus with an albedo of 12$\pm$2\%{}, Fern\'andez et al. 2002) or show temporal cometary activity (like Chiron, with 11$\pm$2\%{}, Groussin et al.2004). The surface layer on Hale--Bopp could have been formed by frost and dust falling back from the coma somewhat before the activity ceased. Ice grains have lower bulk density than dust grains, and they can be dragged away easier than dust. Although in quite different conditions, the scenario of gas jets dragging ice grains was observed in the hyperactive comet 103P/Hartley 2 (A'Hearn et al. 2011). We suspect that in the case of Hale--Bopp, ice grains originated deep from the interior of the nucleus, and were carried up by the sublimating gases, primarily CO. It is also likely that the activity translocated to an interior area of the nucleus, because heat inertia and ice recrystallization were significant drivers of the late activity (Capria et al. 2002). When the comet approaches the Sun, the icy will be dragged away soon, due to the increasing activity. Therefore, had there been a similar layer in the previous approach to the Sun, it should have been removed early by the regular activity, leading to the usual dark surface of an ordinary nucleus.

In summary, Hale-Bopp gave us two major lessons. First, it defeated the paradigm that cometary activity ceases at around 3-5 AU, maintaining its active state by an order of magnitude farther than usual. Second, Hale-Bopp demonstrated that the cessation of cometary activity, at least at large solar distances, can be a much different process from what happens in ordinary comets closer to the Sun: the late activity can effectively restructure the surface of the nucleus. Consequently, giant extrasolar comets may also be very much different objects than the comets we know, with processes specific to giant comets that need to be fully understood for a proper interpretation of cometary activity around other stars.

\section*{Acknowledgements}
This project was supported by the Hungarian OTKA grants K76816, K83790, K104607, the  HUMAN MB08C 81013 project of the MAG Zrt., the PECS-98073 program of the European Space Agency (ESA) and the Hungarian Space Office, the Lend\"ulet 2009/2012 Young Researchers' Programs and the Bolyai Research Fellowship of the Hungarian Academy of Sciences, and the European Community's Seventh Framework Programme (FP7/2007- 2013) under grant agreement no. 269194.

\end{document}